\documentclass[
  aps,
  pre,
  reprint,
  superscriptaddress,
  longbibliography,
  floatfix
]{revtex4-2}

\usepackage{amsmath}
\usepackage{amssymb}
\usepackage{bm}
\usepackage{graphicx}
\usepackage{comment}

\begin{document}

\title{Ordinary Disordered Materials Can Carry Hyperuniform Physical Fields}

\author{Liyu Zhong}
\email{lz8309@princeton.edu}
\affiliation{Department of Mechanical and Aerospace Engineering,
Princeton University, Princeton, New Jersey 08544, USA}

\author{Haina Wang}
\email{hainaw@sas.upenn.edu}
\affiliation{Department of Physics and Astronomy,
University of Pennsylvania, Philadelphia, PA 19104}

\author{Yang Jiao}
\email{yang.jiao.2@asu.edu}
\affiliation{Materials Science and Engineering,
Arizona State University, Tempe, AZ 85287}
\affiliation{Department of Physics,
Arizona State University, Tempe, AZ 85287}

\date{\today}

\begin{abstract} Fluctuations in disordered matter play a central role in determining material properties and physical responses. Recent studies have identified an exotic class of systems known as structurally hyperuniform materials, in which large-scale density fluctuations are anomalously suppressed through special spatial organization of particles, phases, or microstructural features. %Here we show that hyperuniformity can arise through an entirely different mechanism. 
Here we demonstrate that ordinary, structurally nonhyperuniform disordered materials can nevertheless support hyperuniform physical scalar, vector, and tensor fields such as charge, bound current, vorticity, defect density, and stress. We develop a general theoretical framework in which a physical field is generated from a more primitive parent field through a local physical operator. In Fourier space, the spectrum of the derived field is determined by the product of the parent-field spectrum and the Fourier symbol of the operator. When the operator embodies a local gauge-like constraint, its Fourier symbol possesses zeros at small wavenumber, eliminating the corresponding long-wavelength fluctuations. As a consequence, the derived field exhibits complete suppression of infinite-wavelength intensity fluctuations, irrespective of the large-scale disorder and nonhyperuniformity of the parent field. We demonstrate this mechanism in elastic, electrostatic, and magnetostatic settings, showing that operator-generated incompatibility, bound charge, and bound-current fields can become hyperuniform even when their parent eigenstrain, polarization, or magnetization fields remain conventionally disordered. %we show that random eigenstrain fields in elastic solids generate hyperuniform incompatibility fields, which regularize residual-stress fluctuations and eliminate the long-wavelength stress divergences produced by structureless disorder. 
These findings broaden the notion of hyperuniformity from a structural property of matter to a universal field phenomenon generated by local physical constraints. \end{abstract}


\maketitle

\section{Introduction}

Disordered materials exhibit fluctuations over a wide range of length scales, and these fluctuations often govern macroscopic physical behavior. For example, electrical transport, mechanical response, diffusion, wave propagation, and collective dynamics are influenced not only by the average properties of a material but also by the spatial organization of its disorder \cite{torquato2002random,sahimi2003heterogeneous,kirkpatrick1973percolation,sethna2001crackling,ma2023theory}. In many systems, long-wavelength fluctuations play a particularly important role because they persist over large distances and can dominate effective material properties, emergent collective behavior, and system-wide responses. Understanding how such fluctuations of structure and physical fields arise, propagate, and are suppressed is therefore a central problem across condensed-matter physics, materials science, soft matter, and biological systems.

The concept of hyperuniformity identifies a remarkable exception to ordinary disorder, i.e., disordered hyperuniform systems possess a form of hidden long-range order in which large-scale fluctuations are anomalously suppressed \cite{torquato2003local, To18a}. For point patterns, this means that the static structure factor vanishes in the zero-wavenumber limit $\lim_{|{\bf k}|\rightarrow0}S({\bf k})=0$ \cite{torquato2003local}; for heterogeneous media, the analogous quantity is the spectral density of a phase-indicator field \cite{zachary2009hyperuniformity}. The concept of hyperuniformity has since been generalized to scalar, vector and tensor fields \cite{torquato2016generalizations}, weighted particle systems \cite{torquato2026hyperuniformity}, and is now recognized as a unifying principle for a wide class of ordered, disordered, equilibrium, and nonequilibrium systems \cite{To18a}, see {\it Supporting Information} for detailed definitions. Hyperuniform organization has been observed or engineered in diverse settings, including quantum matter \cite{De16, le2017enhanced, Ge19, Ru19, Zh20, sakai2022quantum, chen2025anomalous}, photonic  materials  \cite{Fl09, Ha13, Fl13,man2013isotropic, Mu17, li2018biological}, jammed packings \cite{donev2005unexpected,zachary2011hyperuniform,dale2022hyperuniform}, active fluids \cite{Le19, zhang2022hyperuniform, lei2023does, backofen2024nonequilibrium, leoni2025confinement, maire2026hyperuniformity,huang2021circular}, random organizing systems \cite{Ja15, hexner2017enhanced, lei2019hydrodynamics, wilken2022random}, and biological systems \cite{jiao2014avian,Ma15, ge2023hidden, liu2024universal, li2024fluidization}, to name but a few. Recent work has shown that conservation laws combined with nonequilibrium driving can generically produce structural hyperuniformity, with tunable scaling $S(k)\sim k^m$ controlled by the number of conserved mass multipole moments \cite{maire2025hyperuniformity}. These developments have established the prevailing viewpoint that hyperuniformity is an exotic structural property of the spatial arrangement of particles, phases, or microstructural features in disordered materials.

%\bf we need to provide a thorough introduction to the concept of hyperuniformity, and its generalization to random fields. I will share some tex on this.

Yet many physical responses are not directly controlled by structural, i.e., density or phase-volume fluctuations, they are also governed by physical source fields such as charge, current, force, vorticity, defect density, incompatibility, and stress. 
Although these fields are carried by disordered materials, they are not determined solely by the spatial organization of the underlying structure. 
Instead, they are often generated from more primitive parent fields through local physical laws and constraints. 
For example, bound charge is the divergence of polarization \cite{jackson1999classical,landau1984electrodynamics}; bound current is the curl of magnetization \cite{jackson1999classical,landau1984electrodynamics}; active force is generated from gradients or divergences of active stress \cite{ramaswamy2010mechanics,marchetti2013hydrodynamics,doostmohammadi2018active,alert2022active}; dislocation density arises from gradients of plastic distortion \cite{nye1953some,bilby1955continuous,kroner1981continuum,acharya2001model,groger2008defect}; and incompatibility is generated from eigenstrain in elasticity \cite{eshelby1957determination,mura1987micromechanics}. Many of these field-generating relations originate from local conservation laws, compatibility conditions, and gauge-like constraints, which can selectively remove specific long-wavelength modes, potentially creating a hidden large-scale ``quietness'' that is not apparent in the embedding disordered material. This observation raises a fundamental question: {\it must hyperuniformity of physical fields originate from hyperuniform structure, or can it emerge directly from the physical laws that generate the observable fields?}

%\bf It might be good to show some eye-catching images of the mentioned example fields, if we have space

Here, we answer this question by formulating hyperuniformity as a property of physical-field architecture. Let a scalar, vector, or tensor observable be generated from a parent field by a local physical operator. In the Fourier space, the spectrum of the derived field is controlled by both the spectrum of the parent field and the Fourier symbol of the operator. When the operator symbol vanishes at small wavenumber, the corresponding long-wavelength components of the derived field are eliminated. Consequently, even a parent field with ordinary disorder and nonzero infinite-wavelength fluctuations can generate a hyperuniform physical source via the local operator. This mechanism is distinct from designing or discovering a hyperuniform structure directly, i.e., the hidden order need not reside in where matter is located, but can emerge from the local physical laws that generate the fields.

To demonstrate principles of this this new framework of operator-induced hyperuniformity, we first show an example of a random eigenstrain or plastic-strain field in a disordered elastic solid. 
The parent eigenstrain may represent local transformation strain, plastic rearrangement, growth strain, or shear-transformation-like activity, and is structurally ordinary in the sense that its long-wavelength spectrum is not suppressed. Nevertheless, the geometrical incompatibility generated from this tensor parent field is hyperuniform at large scales, characterized by small-$k$ scaling $\sim k^4$ of the spectral density. 
In two dimensions, incompatibility is a scalar source produced by a second-order compatibility operator acting on the tensor eigenstrain field. 
Thus, this example is not a claim that the full tensor eigenstrain spectral matrix is itself hyperuniform; 
rather, it shows that a physically meaningful scalar source coupled to stress can become hyperuniform through tensorial geometry.

This source-field derived hyperuniformity has direct physical consequences. 
A nonhyperuniform incompatibility source with long-wavelength fluctuations can be strongly amplified by the elastic Green function, whereas an eigenstrain-generated incompatibility source of the same overall strength carries operator-imprinted zeros at small wavenumber and is hyperuniform. The elastic response operator is identical in both cases; yet the large-scale fluctuations of the source fields can lead to distinctly different responses: The hyperuniform incompatibility produces a regular residual-stress spectrum, whereas a nonhyperuniform source produces a long-wavelength stress catastrophe. %The material response is therefore largely controlled by whether the source inherits the large-scale architecture imposed by the physical operator that generated it.

In addition to this second-order incompatibility example, we demonstrate two classes of first-order operators that produce hyperuniform source field. 
In electrostatics, an ordinary disordered polarization field generates bound charge through the divergence operator, $\rho_b=-\nabla\cdot\mathbf P$, imprinting a $k^2$ low-wavenumber source spectral density that compensates the Poisson field-response singularity and regularizes the electric-field spectrum. %relative to a matched white-charge source. 
In the magnetostatic analogue, an ordinary in-plane magnetization field generates an out-of-plane bound current through the curl operator, $J_b=(\nabla\times\mathbf M)_z$, again producing a $k^2$ source spectral density and a regular magnetic-field response relative to a matched white-current source. 
Together, the elastic, electrostatic, and magnetostatic examples show that operator-induced hyperuniformity is not a peculiarity of one material model, but a common response-level fingerprint of local physical constraints.

This work broadens hyperuniformity from a property of material arrangement to a property of physical-field generation. Structural hyperuniformity asks whether density or phase fluctuations vanish at large scales. Field hyperuniformity asks whether the physical sources that govern response are quiet, or have vanishing signal, at large scales, even when the underlying material structure or parent field is not. The relevant object may be %a scalar source generated from a tensor parent field, a charge generated from a vector polarization, a current generated from a magnetization, or more generally 
any source generated through a local operator with a nontrivial null space. 
Thus, the material response is controlled not only by the nature of disorder of the underlying structure, but also by the competition between source-operator zeros and Green-function singularities.

\section{Results}

\subsection{Physical operators turn noisy parent fields into hyperuniform sources}

%The mechanism of field hyperuniformity is simple but physically restrictive: many sources that enter continuum response equations are not arbitrary random fields.  They are obtained by differentiating a more primitive parent field. A derivative removes spatially uniform modes.  In Fourier space, this means that every derivative contributes a factor of $k$; therefore a source produced by a first-order operator loses its lowest long-wavelength modes, while a source produced by a second-order operator loses them even more strongly.  This is the origin of the quietness discussed below.

The mechanism of field hyperuniformity is simple but physically restrictive: many sources that enter continuum response equations are not arbitrary random fields, but are generated from more primitive parent fields through local gauge-like constraints. 
These constraints typically act through differential operators, and therefore eliminate spatially uniform modes. In Fourier space, each derivative contributes a factor of $k$, so a source generated by a first-order operator has its lowest long-wavelength components suppressed, while a source generated by a second-order operator suppresses them even more strongly. Thus, the large-scale quietness discussed below is not imposed by structural order in the parent field, but emerges kinematically from the gauge-like constraint that defines the physical source.

Let $\psi(\mathbf{x})$ be a parent field and let the physical source be
\begin{equation}
q(\mathbf{x})=A(\nabla)\psi(\mathbf{x}),
\label{eq:operator_source}
\end{equation}
where $A(\nabla)$ is a local field-generating operator, often arising from a conservation law, compatibility condition, or gauge-like constraint.
In Fourier space,
\begin{equation}
\widehat q(\mathbf{k})=A(\mathbf{k})\widehat\psi(\mathbf{k}).
\end{equation}
If the operator has a low-wavenumber zero,
\begin{equation}
A(\mathbf{k})\sim k^m A_m(\widehat{\mathbf{k}}),
\label{eq:operator_order}
\end{equation}
then the spectrum of the derived field is multiplied by the squared operator
symbol:
\begin{equation}
{\tilde{\chi}}_q(\mathbf{k})
\sim
k^{2m} A_m(\widehat{\mathbf{k}})\tilde{\chi}_\psi(\mathbf{k})A_m^\dagger(\widehat{\mathbf{k}}).
\label{eq:operator_induced_suppression}
\end{equation}
\noindent Thus, if the parent field is an ordinary short-range disordered field with
$\tilde{\chi}_\psi(k)\sim k^0$, a first-order source has
$\tilde{\chi}_q(k)\sim k^2$, and a second-order source has
$\tilde{\chi}_q(k)\sim k^4$. The material can therefore be noisy, i.e.,
nonhyperuniform, in the parent field but hyperuniform in the physical source
field that couples to a response equation. More generally, if the source then
drives a response through a Green operator with an infrared pole, the observable
response is controlled by a competition between the operator-imprinted source
zero and the Green-function singularity. This source-response viewpoint is the
central organizing principle of the numerical examples below.

This mechanism is not an abstract construction; it appears in many familiar
physical laws. A polarized medium carries bound charge
\begin{equation}
\rho_b=-\nabla\cdot \mathbf{P}.
\label{eq:bound_charge}
\end{equation}
A spatially uniform polarization does not create charge; only spatial
variation does. Therefore, for a short-range disordered polarization field,
\begin{equation}
\widehat\rho_b(\mathbf{k})=-i k_i \widehat P_i(\mathbf{k}),
\qquad
\tilde{\chi}_{\rho_b}(k)\sim k^2.
\end{equation}
Thus, the divergence operator converts noisy polarization into a charge source
with suppressed large-scale fluctuations.

The same structure appears in active matter. The active force density is
generated from active stress by
\begin{equation}
f_i=\partial_j\sigma^a_{ij}.
\label{eq:active_force}
\end{equation}
A uniform active stress does not act as a body force; only gradients of active
stress do. Hence,
\begin{equation}
\widehat f_i(\mathbf{k})=i k_j\widehat\sigma^a_{ij}(\mathbf{k}),
\qquad
\tilde{\chi}_f(k)\sim k^2
\end{equation}
when the parent active stress possesses ordinary short-range correlation. The
physical source that drives mechanical motion is therefore ``quieter'' at long
wavelengths, i.e., hyperuniform, than the active stress field from which it is
generated.

Curl-type operators have the same consequence. Vorticity is generated from
velocity by
\begin{equation}
\boldsymbol{\omega}=\nabla\times\mathbf{v},
\end{equation}
so uniform translation carries no vorticity. Similarly, an in-plane
magnetization field $\mathbf{M}=(M_x,M_y)$ generates an out-of-plane bound
current
\begin{equation}
J_b=(\nabla\times\mathbf{M})_z
=
\partial_x M_y-\partial_y M_x,
\label{eq:bound_current}
\end{equation}
and therefore
\begin{equation}
\widehat J_b(\mathbf{k})
=
i k_x \widehat M_y(\mathbf{k})
-
i k_y \widehat M_x(\mathbf{k}),
\qquad
\tilde{\chi}_{J_b}(k)\sim k^2
\end{equation}
for an ordinary short-range disordered magnetization parent field. In defect
mechanics, the Nye dislocation-density tensor is generated from the plastic
distortion by
\begin{equation}
\alpha_{ij}=\epsilon_{jkl}\partial_k\beta^p_{il}.
\label{eq:nye_tensor}
\end{equation}
Again, the derivative removes uniform plastic-distortion modes and imposes a
low-$k$ suppression on the derived defect-density field. These examples show
that scalar, vector, and tensor physical sources can all inherit the strong
suppression of long-wavelength fluctuations from the differential operators
that define them.

The strongest suppression arises when the relevant physical source is produced
by a second-order geometrical operator. In an elastic solid with eigenstrain
or plastic strain, the incompatibility source is not an independent random
field. In two dimensions it is
\begin{equation}
\eta=
\mathrm{inc}\,\varepsilon^*
=
\partial_{xx}\varepsilon^*_{yy}
+
\partial_{yy}\varepsilon^*_{xx}
-
2\partial_{xy}\varepsilon^*_{xy}.
\label{eq:incompatibility_2d}
\end{equation}
Equivalently,
\begin{equation}
\widehat\eta(\mathbf{k})
=
-k_y^2\widehat\varepsilon^*_{xx}
+
2k_xk_y\widehat\varepsilon^*_{xy}
-
k_x^2\widehat\varepsilon^*_{yy}.
\label{eq:incompatibility_fourier}
\end{equation}
Every term contains two powers of $k$. Therefore, if the parent eigenstrain
field is a conventional short-range disordered tensor field, the scalar
incompatibility source satisfies
\begin{equation}
\tilde{\chi}_\eta(k)\sim k^4.
\label{eq:incompatibility_k4}
\end{equation}
Thus, the central physical mechanism is that ordinary random eigenstrain can generate a hyperuniform source, not because the parent field is hyperuniform, but because the incompatibility field is generated through a compatibility constraint that acts as a gauge-like operator. Here ``gauge-like'' is used in a broad sense to describe conservation laws, compatibility conditions, and differential constraints in which the physical source depends only on equivalence classes of a parent field. %Additions lying in the null space of the source-generating operator do not change the source. These operators encode a redundancy between parent fields and observable sources; 
The null spaces of these operators remove uniform or otherwise unobservable modes and impose the long-wavelength constraints responsible for field hyperuniformity. %The incompatibility operator possesses a nontrivial null space and suppresses long-wavelength source modes, imprinting zeros in the spectrum of the derived field. 

%Here ``gauge-like'' is used in a broad sense to describe conservation laws, compatibility conditions, and differential constraints in which the physical source depends only on equivalence classes of a parent field. Additions lying in the null space of the source-generating operator do not change the source. Thus, a uniform polarization produces no bound charge, a uniform active stress produces no body force, a uniform velocity produces no vorticity, a uniform or compatible plastic distortion produces no dislocation density, and compatible eigenstrain produces no incompatibility. 

Operator-induced hyperuniformity is therefore a statement about the architecture of the physical source rather than the disorder of the parent field alone. The relevant question is not simply whether the material structure or parent field is hyperuniform, but whether the source field that enters the governing response equations retains any long-wavelength fluctuations after the action of the underlying physical constraints. In the following sections, we show concrete examples of this principle in elastic, electrostatic, and magnetostatic problems: the parent field exhibits ordinary short-range disorder, whereas the derived source field is strongly hyperuniform due to the gauge-like compatibility constraint that generates it.

%This is the central physical mechanism used in the following sections: ordinary random eigenstrain can generate a hyperuniform source, not because the parent field is hyperuniform, but because the incompatibility operator suppresses its long-wavelength source modes. 

%Field hyperuniformity is therefore a statement about the architecture of the physical source.  The relevant question is not only whether the material structure or parent field is noisy, but whether the source field that enters a response equation retains those long-wavelength fluctuations.  In the next section we show that random eigenstrain inclusions provide a concrete example: the parent tensor field is ordinary, whereas the derived scalar incompatibility source is strongly quiet at large scales.

\subsection{Random eigenstrain inclusions are noisy in strain but hyperuniform in incompatibility}

%We first test the operator-induced mechanism in a minimal elastic setting.  
We construct random local eigenstrain inclusions in a two-dimensional periodic
disordered solid. 
Here $L$ is the box side length and $N$ is the grid resolution, not the number of inclusions; we use $\Delta x=L/N=1$, so they are numerically equal in the simulations. 
The parent field is the tensor eigenstrain
$\varepsilon^*(\mathbf{x})$, which represents local transformation strain,
plastic strain, or shear-like rearrangement. 
The inclusions are placed without
imposing any hyperuniform constraint on their centers or amplitudes. Thus the
parent eigenstrain is an ordinary short-range disordered field: its
low-wavenumber spectral density is not suppressed.

\begin{figure*}[t!]
\centering
\includegraphics[width=\textwidth]{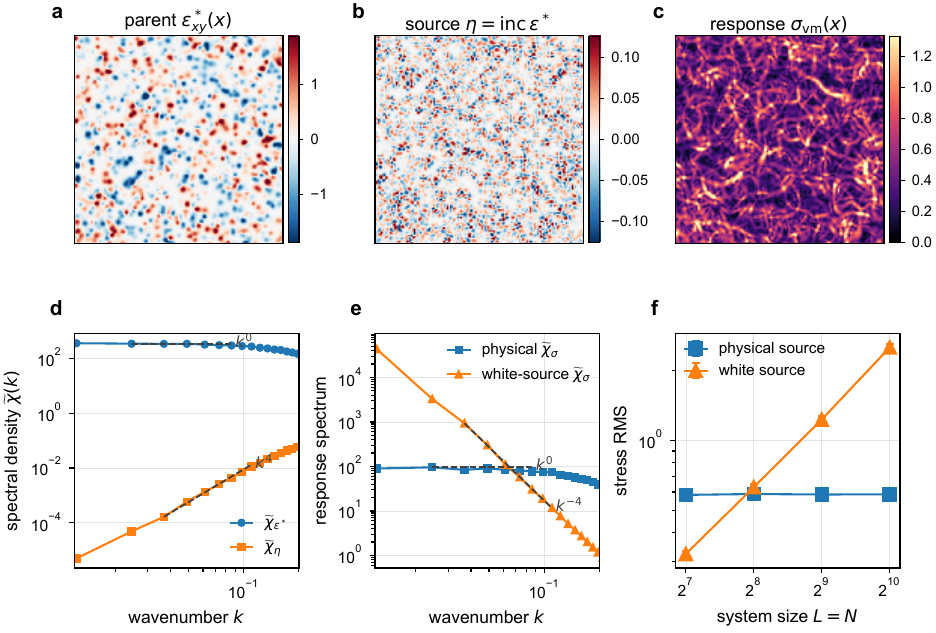}
\caption{\textbf{Operator-induced incompatibility hyperuniformity and residual-stress regularization.}
(A) Representative shear component of the parent eigenstrain field,
$\varepsilon^*_{xy}(\mathbf{x})$. The field is disordered and has no imposed
large-scale hyperuniform constraint. (B) The scalar incompatibility source
$\eta=\operatorname{inc}\varepsilon^*$ derived from the same eigenstrain field.
The second-order compatibility operator produces a finer source pattern and
removes long-wavelength modes. (C) Residual von Mises stress generated by the
physical incompatibility source. (D) Radially averaged spectral densities of
the parent eigenstrain and the derived incompatibility. The parent spectrum has
an ordinary low-$k$ plateau, $\tilde{\chi}_{\varepsilon^*}(k)\sim k^0$, whereas
the derived incompatibility follows the second-order-operator prediction
$\tilde{\chi}_{\eta}(k)\sim k^4$. (E) Residual-stress spectra produced by the
physical incompatibility source and by an rms-matched white incompatibility
source. The physical source gives a regular low-$k$ stress spectrum,
$\tilde{\chi}_{\sigma}(k)\sim k^0$, whereas the white source produces a
long-wavelength divergence, $\tilde{\chi}_{\sigma}(k)\sim k^{-4}$. (F)
Finite-size stress rms. The physical-source response remains approximately
size independent, whereas the white-source response grows strongly with system
size. Spectral panels use $N=512$ with $\Delta x=1$; finite-size results use
$N=128,256,512,1024$.}
\label{fig:incompatibility_response}
\end{figure*}

Figure~\ref{fig:incompatibility_response} shows a representative realization
and the corresponding spectra. The parent shear eigenstrain component is
visually noisy and spatially nonhyperuniform
(Fig.~\ref{fig:incompatibility_response}A). The incompatibility obtained from
the same field is also spatially disordered, but it is generated by the
second-order compatibility operator
(Fig.~\ref{fig:incompatibility_response}B). In Fourier space, this operator
multiplies the parent eigenstrain by two powers of $k$, and therefore its
spectral density must acquire four powers of $k$ at small wave number. The spectra confirm this prediction
(Fig.~\ref{fig:incompatibility_response}D). At the $N=512$ grid resolution, the fitted low-$k$
exponent of the parent eigenstrain spectrum is
$\beta=-0.038\pm0.033$, close to the ordinary-disorder value $\beta=0$. By
contrast, the incompatibility spectrum has exponent
$\beta=3.975\pm0.066$, consistent with the predicted
$\tilde{\chi}_{\eta}(k)\sim k^4$ scaling. At $N=1024$, the source exponent
remains close to the same asymptotic value,
$\beta=3.893\pm0.042$. These results show that hyperuniformity is not a
property of the parent eigenstrain itself, and is created when the physical
source field is formed.

%This distinction is important because $\eta$ is the source that enters the elastic compatibility problem. 
These results indicate although the eigenstrain field is tensorial and nonhyperuniform,
the scalar incompatibility derived from it is a hyperuniform field. The example is therefore not a conventional structural-hyperuniform
medium, nor is it a claim that the full tensor eigenstrain spectral matrix is
hyperuniform. Rather, it is a physically generated scalar source whose
long-wavelength modes are suppressed by tensorial geometry. The material is
{\it structurally nonhyperuniform} in the parent field but {\it physically hyperuniform} in the source
field that drives residual stress.

\subsection{Source field hyperuniformity regularizes the residual-stress response}

The hyperuniformity of incompatibility has a direct physical consequence.
In two-dimensional isotropic elasticity, the residual stress generated by an
incompatibility source can be represented through an Airy stress function
$\chi$,
\begin{equation}
\sigma_{xx}=\partial_{yy}\chi,\qquad
\sigma_{yy}=\partial_{xx}\chi,\qquad
\sigma_{xy}=-\partial_{xy}\chi,
\label{eq:airy_stress}
\end{equation}
with the compatibility equation
\begin{equation}
\Delta^2 \chi = Y\eta,
\label{eq:biharmonic_response}
\end{equation}
where $Y$ is an elastic modulus. Constants and tensorial angular projectors do
not affect the low-$k$ power counting. Since
Eq.~\ref{eq:biharmonic_response} contains the inverse biharmonic operator,
while Eq.~\ref{eq:airy_stress} contains two derivatives, the stress response to
incompatibility scales as
\begin{equation}
\widehat{\sigma}(\mathbf{k})\sim
{\widehat{\eta}(\mathbf{k})}/{k^2},
\qquad
\tilde{\chi}_{\sigma}(k)
\sim
{\tilde{\chi}_{\eta}(k)}/{k^4}.
\label{eq:stress_filter}
\end{equation}
Thus, the same elastic solid responds very differently to two sources of the
same overall strength but different long-wavelength architecture. A physical
source generated from eigenstrain has $\tilde{\chi}_{\eta}(k)\sim k^4$, so its
stress spectrum is regular, $\tilde{\chi}_{\sigma}(k)\sim k^0$. A
structureless white incompatibility source has $\tilde{\chi}_{\eta}(k)\sim
k^0$, so the same elastic response produces
$\tilde{\chi}_{\sigma}(k)\sim k^{-4}$.

Figure~\ref{fig:incompatibility_response}E verifies this response-level
prediction. For the physical incompatibility generated from eigenstrain, the
residual-stress spectrum remains finite as $k\to0$. At $N=512$, the fitted
stress exponent is $\beta=-0.022\pm0.053$, consistent with the predicted
plateau. For the matched white source, the stress spectrum instead grows
rapidly at small wave number, with fitted exponent
$\beta=-4.005\pm0.043$, in agreement with the $k^{-4}$ elastic amplification.
At $N=1024$, the physical response remains close to a plateau
$(\beta=-0.044\pm0.041)$, while the white-source response remains close to the
asymptotic $k^{-4}$ law $(\beta=-3.994\pm0.027)$. The contrast is not caused by
a difference in total source strength: in the numerical comparison, the white
incompatibility source is normalized to have the same rms amplitude as the
physical incompatibility source.

The same distinction appears in the finite-size stress amplitude
(Fig.~\ref{fig:incompatibility_response}F). Across
$N=128,256,512,1024$, the physical-source stress rms remains approximately
constant, near $0.58$, showing no systematic growth with system size. By
contrast, the matched white-source stress rms grows from $0.324$ to $2.512$,
an increase by a factor of approximately $7.74$. This finite-size behavior
follows directly from the spectral powers. In two dimensions, a regular stress
spectrum gives a finite large-scale contribution to
$\langle\sigma^2\rangle$, whereas a $k^{-4}$ stress spectrum gives
$\langle\sigma^2\rangle\sim L^2$ and hence
$\sigma_{\mathrm{rms}}\sim L$.

%The response calculation reveals why field hyperuniformity is physically observable. 

We note that the elastic Green function is identical in the two comparisons; only the source field is changed. An arbitrary nonhyperuniform source contains long-wavelength fluctuations that are strongly amplified by the inverse elastic operator, leading to a stress catastrophe. In contrast, a source generated through the appropriate compatibility constraint carries operator-imprinted zeros that remove these long-wavelength modes and exactly cancel the Green-function singularity. The resulting residual-stress spectrum remains finite and well behaved. Thus, material response is governed not only by the amount of disorder present, but also by the large-scale architecture of the physical source field and the gauge-like constraints that generate it.

%The response calculation shows why field hyperuniformity is physically observable.  The elastic Green function is identical in the two comparisons; only the source architecture is changed.  If the source is treated as an arbitrary white random field, the inverse elastic operator creates a long-wavelength stress catastrophe.  If the source is generated by the correct physical operator from a noisy parent field, the operator-imprinted low-$k$ zeros exactly cancel the Green-function singularity.  The residual-stress response is therefore controlled not only by the variance of the disorder, but by whether the source carries the long-wavelength constraints imposed by its physical origin.

%\bf Is it possible to keep the detailed discussions in SI, but show the key figures for these two additional systems? We could combine current Fig. 1 and Fig. 2, then plan for new figure for change density and another figure for active force?

\subsection{First-order operator families leading to hyperuniform electrostatic and magnetostatic responses}

The aforementioned elastic example involves a second-order incompatibility operator. We next show that 
two independent first-order operator families, divergence and curl, produce the
same $k^2$ source hyperuniformity and regularize the corresponding field
responses, further demonstrating
that the mechanism identified above is an operator-level principle shared by familiar physical
constraints. The discussions are kept simple and more details are given in {\it Supporting Information}.
%These examples are deliberately simple. Their role is not to introduce additional microscopic models, but to show that the mechanism identified above is an operator-level principle shared by familiar physical constraints.

\begin{figure*}[t!]
\centering
\includegraphics[width=\textwidth]{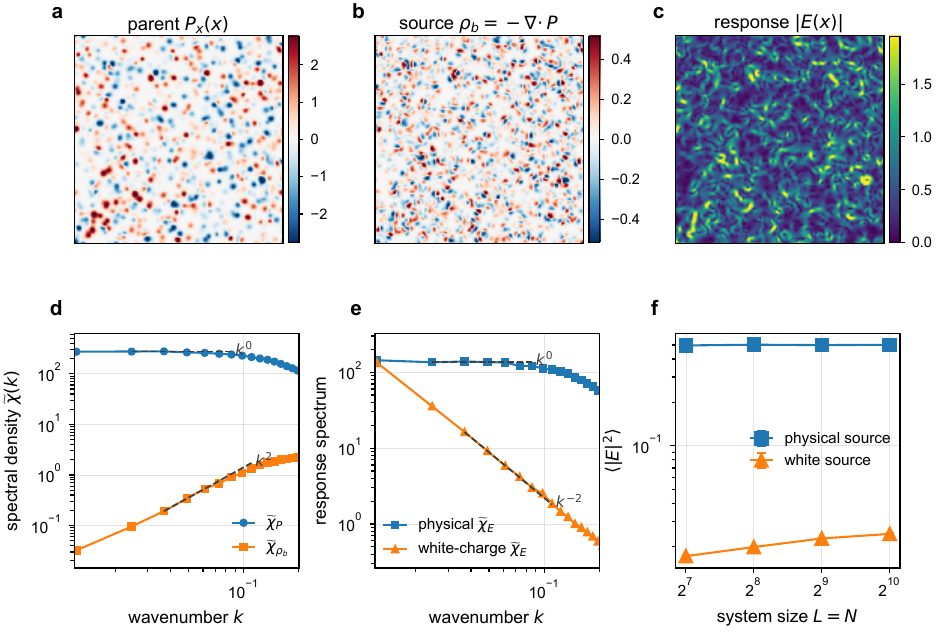}
\caption{\textbf{A first-order divergence operator generates quiet bound charge and regularizes the electric field.}
(A) Representative component of the parent polarization field,
$P_x(\mathbf{x})$. (B) Bound charge
$\rho_b=-\nabla\cdot\mathbf{P}$ generated from the same polarization field.
(C) Magnitude of the electric-field response, $|\mathbf{E}(\mathbf{x})|$.
(D) Radially averaged spectral densities of the parent polarization and bound
charge. The parent field is ordinary, with
$\tilde{\chi}_{P}(k)\sim k^0$, whereas the divergence-generated source follows
$\tilde{\chi}_{\rho_b}(k)\sim k^2$. (E) Electric-field spectra generated by the
physical bound charge and by an rms-matched white charge source. The physical
source gives a regular low-$k$ electric-field spectrum,
$\tilde{\chi}_{E}(k)\sim k^0$, whereas the white charge source produces the
Poisson infrared enhancement, $\tilde{\chi}_{E}(k)\sim k^{-2}$. (F)
Finite-size electric-field amplitude. The physical-source response remains
approximately size independent, while the white-source control grows weakly, as
expected for the two-dimensional Poisson problem. Spectral panels use
$N=512$ with $\Delta x=1$; finite-size results use $N=128,256,512,1024$.}
\label{fig:bound_charge_response}
\end{figure*}

For the divergence family, a polarization field generates the bound charge
\begin{equation}
\rho_b=-\nabla\cdot\mathbf{P},
\qquad
\widehat{\rho}_b(\mathbf{k})=-i k_i\widehat P_i(\mathbf{k}).
\label{eq:divergence_bound_charge_results}
\end{equation}
Thus, for an ordinary short-range disordered polarization parent field,
\begin{equation}
\tilde{\chi}_{\rho_b}(k)\sim k^2.
\label{eq:bound_charge_k2_results}
\end{equation}
The electrostatic response is determined by the periodic Poisson problem
\begin{equation}
-\Delta\phi=\rho_b,\qquad
\mathbf{E}=-\nabla\phi.
\label{eq:poisson_response_results}
\end{equation}
In Fourier space, $\widehat{\phi}(\mathbf{k})=\widehat{\rho}_b(\mathbf{k})/k^2$
and $\widehat{E}_i(\mathbf{k})=-i k_i\widehat{\phi}(\mathbf{k})$, so
\begin{equation}
\tilde{\chi}_{E}(k)
\sim
{\tilde{\chi}_{\rho_b}(k)}/{k^2}.
\label{eq:electric_filter_results}
\end{equation}
A physical bound-charge source therefore yields
$\tilde{\chi}_{E}(k)\sim k^0$, whereas an rms-matched white charge source
yields $\tilde{\chi}_{E}(k)\sim k^{-2}$.

Figure~\ref{fig:bound_charge_response} confirms this prediction. At
$N=512$, the bound-charge source exponent is
$\beta=1.883\pm0.062$, already close to the expected value $2$, while the
physical electric-field response has exponent
$\beta=-0.009\pm0.066$, consistent with a regular plateau. The white-charge
control gives $\beta=-1.972\pm0.074$, close to the predicted $k^{-2}$ infrared
enhancement. At $N=1024$, the asymptotic scaling becomes even clearer: the
source exponent is $\beta=1.970\pm0.049$, the physical electric response is
$\beta=0.001\pm0.046$, and the white-source response is
$\beta=-1.989\pm0.014$. The finite-size contrast is weaker than in the elastic
case because the two-dimensional Poisson response to white charge produces only
a weak large-scale growth in the integrated field amplitude, but the spectral
distinction between the physical source and the white-source control is
unambiguous.

\begin{figure*}[t!]
\centering
\includegraphics[width=\textwidth]{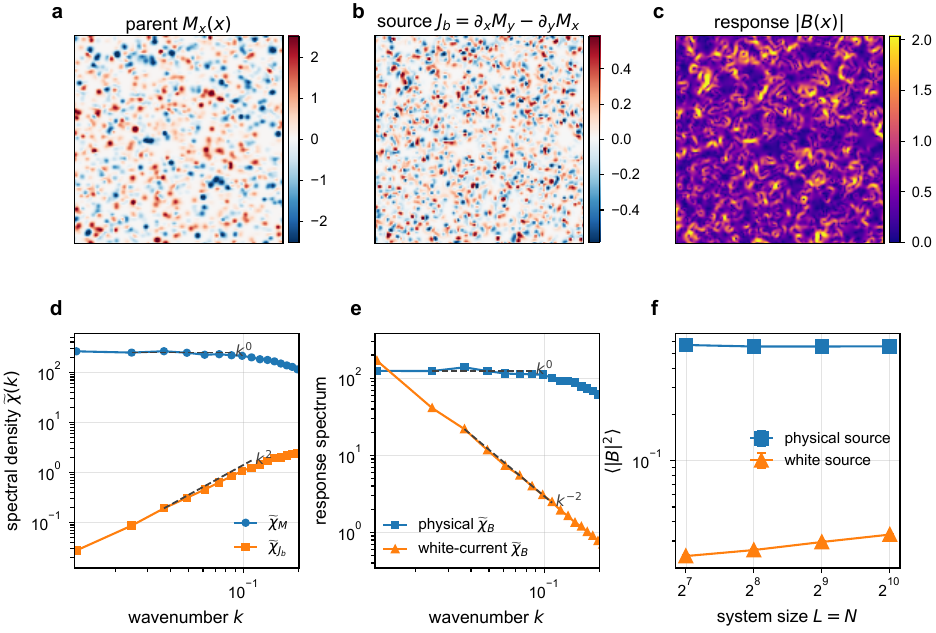}
\caption{\textbf{A first-order curl operator generates quiet bound current and regularizes the magnetic field.}
(A) Representative component of the parent magnetization field,
$M_x(\mathbf{x})$. (B) Out-of-plane bound current
$J_b=(\nabla\times\mathbf{M})_z=\partial_xM_y-\partial_yM_x$ generated from the
same magnetization field. (C) Magnitude of the magnetic-field response,
$|\mathbf{B}(\mathbf{x})|$. (D) Radially averaged spectral densities of the
parent magnetization and bound current. The parent field is ordinary, with
$\tilde{\chi}_{M}(k)\sim k^0$, whereas the curl-generated source follows
$\tilde{\chi}_{J_b}(k)\sim k^2$. (E) Magnetic-field spectra generated by the
physical bound current and by an rms-matched white current source. The physical
source gives a regular low-$k$ magnetic-field spectrum,
$\tilde{\chi}_{B}(k)\sim k^0$, whereas the white current source produces the
magnetostatic infrared enhancement, $\tilde{\chi}_{B}(k)\sim k^{-2}$. (F)
Finite-size magnetic-field amplitude. The physical-source response remains
approximately size independent, while the white-source control grows weakly.
Spectral panels use $N=512$ with $\Delta x=1$; finite-size results use
$N=128,256,512,1024$.}
\label{fig:bound_current_response}
\end{figure*}

For the curl family, an in-plane magnetization
$\mathbf{M}=(M_x,M_y)$ generates an out-of-plane bound current
\begin{equation}
J_b=(\nabla\times\mathbf{M})_z
=
\partial_xM_y-\partial_yM_x,
\end{equation}
\begin{equation}
\widehat{J}_b(\mathbf{k})
=
i k_x\widehat M_y(\mathbf{k})
-
i k_y\widehat M_x(\mathbf{k}).
\label{eq:curl_bound_current_results}
\end{equation}
This is again a first-order source-generating operator. Therefore, for an
ordinary short-range disordered magnetization field,
\begin{equation}
\tilde{\chi}_{J_b}(k)\sim k^2.
\label{eq:bound_current_k2_results}
\end{equation}
The corresponding two-dimensional magnetostatic response can be written in
terms of an out-of-plane vector potential,
\begin{equation}
-\Delta A_z=\mu_0 J_b,
\end{equation}
\begin{equation}
\mathbf{B}=\nabla\times(A_z\mathbf{e}_z)
=
(\partial_y A_z,-\partial_x A_z).
\label{eq:magnetostatic_response_results}
\end{equation}
Thus,
\begin{equation}
\tilde{\chi}_{B}(k)
\sim
{\tilde{\chi}_{J_b}(k)}/{k^2}.
\label{eq:magnetic_filter_results}
\end{equation}
A physical curl-generated current gives
$\tilde{\chi}_{B}(k)\sim k^0$, whereas an rms-matched white current source gives
$\tilde{\chi}_{B}(k)\sim k^{-2}$.

Figure~\ref{fig:bound_current_response} verifies the magnetostatic analogue of
the electrostatic result. At $N=512$, the bound-current source exponent is
$\beta=1.803\pm0.069$, while the physical magnetic response is
$\beta=0.027\pm0.044$ and the white-current control is
$\beta=-2.016\pm0.071$. At $N=1024$, the source exponent reaches
$\beta=1.980\pm0.043$, the physical magnetic response remains regular with
$\beta=-0.007\pm0.033$, and the white-current response gives
$\beta=-2.000\pm0.019$. Thus, the divergence-generated charge and
curl-generated current examples provide two independent first-order
counterparts to the second-order incompatibility example: ordinary parent
fields are converted into hyperuniform physical sources, and those source zeros
compensate the infrared poles of the corresponding Green functions.

\subsection{Gauge-like constraints control large-scale response}

%\bf this section can be re-organized and shortened

The numerical results reveal a simple but powerful chain of consequences. Across the three examples, the response-level fingerprint is determined by the
competition between source zeros and Green-function poles. The second-order
incompatibility operator gives
$\tilde{\chi}_{\eta}(k)\sim k^4$, which cancels the elastic stress
amplification and yields
$\tilde{\chi}_{\sigma}(k)\sim k^0$; a white incompatibility source instead
produces $\tilde{\chi}_{\sigma}(k)\sim k^{-4}$. The first-order divergence and
curl operators give
$\tilde{\chi}_{\rho_b}(k)\sim k^2$ and
$\tilde{\chi}_{J_b}(k)\sim k^2$, which compensate the Poisson and
magnetostatic field responses and yield
$\tilde{\chi}_{E}(k)\sim k^0$ and
$\tilde{\chi}_{B}(k)\sim k^0$; white charge and white current controls instead
produce the corresponding $k^{-2}$ infrared enhancements. The same Green
operators are used in the physical-source and white-source comparisons. What
changes is the long-wavelength architecture of the source entering those
operators.

%Ordinary eigenstrain disorder is first transformed by the incompatibility operator into a hyperuniform source with $\tilde{\chi}_{\eta}(k)\sim k^4$. The subsequent elastic response then converts this source into a regular residual-stress spectrum with $\tilde{\chi}_{\sigma}(k)\sim k^0$. The first step is geometrical: the compatibility constraint, acting as a gauge-like operator, removes the long-wavelength modes of the parent tensor field. The second step is mechanical: the inverse elastic response amplifies low-$k$ source modes through a $k^{-4}$ Green-function singularity. The residual stress remains finite because these two effects exactly balance.

%The numerical results establish a chain of implications: ordinary eigenstrain disorder is transformed by the incompatibility operator into a quiet source with $S_\eta(k)\sim k^4$, and the subsequent elastic response converts this source into a regular residual-stress spectrum with $S_\sigma(k)\sim k^0$. The first step is geometrical: the incompatibility operator removes the long-wavelength modes of the parent tensor field.  The second step is mechanical: the inverse elastic response amplifies low-$k$ source modes by $k^{-4}$ in the stress spectrum.  The physical source is regular because these two operations balance each other.

This mechanism clarifies what is fundamentally new about field hyperuniformity.
Conventional hyperuniformity concerns the structure of matter, as characterized
by density, phase, or point-pattern fluctuations. Here, the parent disorder
remains entirely ordinary, yet the derived field that enters the governing
response equation is hyperuniform, leading to the suppression of large-scale 
fluctuations in material responses, such as the elimination of a stress catastrophe. The controlling
variable is therefore the architecture of the physical source field, and the
gauge-like constraints that generate it, rather than the parent disorder alone.

The same logic applies whenever a response equation is driven by a source
generated through a local gauge-like constraint. Bound charge from
polarization, bound current from magnetization, force from active stress,
vorticity from velocity, dislocation density from plastic distortion, and
incompatibility from eigenstrain are all produced by differential operators
with nontrivial null spaces, which impose long-wavelength constraints on the
derived source fields. Whether these constraints regularize the response
depends on the balance between operator-induced source suppression and the
singularity of the relevant Green function. The three examples shown here
demonstrate this balance for one second-order compatibility operator and two
first-order divergence/curl operators. Additional examples and numerical
details, including fitting windows, shell counts, bootstrap uncertainties, and
finite-size trends, are provided in the \textit{SI Appendix}.

\section{Conclusion and Discussion}

We term this phenomenon \emph{operator-induced hyperuniformity}: the emergence of hyperuniform physical source fields from nonhyperuniform parent fields through local gauge-like constraints encoded in physical operators.
Many physical sources are not independent random fields, but are generated from more primitive parent fields by differential operators whose Fourier symbols vanish at small wavenumber.
These operator-imposed zeros suppress long-wavelength source modes, allowing an ordinary short-range-disordered parent field to generate a hyperuniform source.
This mechanism provides a route to hyperuniformity rooted not in material arrangement, but in physical-field generation.
We demonstrated its generality through three numerical realizations spanning distinct operator families in elasticity, electrostatics, and magnetostatics.

\begin{comment}
In random eigenstrain solids, the parent tensor eigenstrain has an ordinary low-$k$ spectrum, whereas the scalar incompatibility generated from it satisfies
$\tilde{\chi}_{\eta}(k)\sim k^4$. This second-order source zero has a direct
mechanical consequence: under the same elastic Green response, the physical
incompatibility source produces a regular residual-stress spectrum and an
approximately size-independent stress rms, whereas a matched structureless
white incompatibility source produces a $k^{-4}$ stress spectrum and a stress
rms that grows strongly with system size. In electrostatics, a noisy
polarization field generates bound charge
$\rho_b=-\nabla\cdot\mathbf P$ with
$\tilde{\chi}_{\rho_b}(k)\sim k^2$, which compensates the Poisson
field-response singularity and yields a regular electric-field spectrum. In
the magnetostatic analogue, an in-plane noisy magnetization field generates
out-of-plane bound current
$J_b=(\nabla\times\mathbf M)_z$ with
$\tilde{\chi}_{J_b}(k)\sim k^2$, again producing a regular magnetic-field
response relative to a matched white-current control. Thus, response is
controlled not only by disorder amplitude, but by the long-wavelength
architecture of the source.
\end{comment}

Our work re-frames hyperuniformity as a property of physical sources rather than
structure alone. Structural hyperuniformity asks whether density or phase
fluctuations vanish at large scales; field hyperuniformity asks whether the
sources entering response equations are quiet at large scales. A material can
therefore be structurally ordinary in its parent disorder while being
hyperuniform in charge, current, force, vorticity, defect density,
incompatibility, or another derived source field. 
%In the elastic example, the hyperuniform object is the scalar incompatibility source generated from tensor eigenstrain, not necessarily the full tensor eigenstrain spectral matrix itself.
Response regularization occurs only when the small-$k$
zeros of the source-generating operator are strong enough to compensate the
singularity of the response kernel.

The broader implication is that long-wavelength response is governed by a
competition between source architecture and Green-function amplification.
White sources probe response singularities directly and can produce
large-scale catastrophes, whereas physically generated sources may contain
hidden cancellations that remove or weaken them. In the examples studied here,
a second-order incompatibility zero cancels the elastic stress amplification,
while first-order divergence and curl zeros cancel the corresponding
electrostatic and magnetostatic field-response poles. Other physical systems,
including active stresses, plastic distortions, defect densities, and
hydrodynamic forcing, can be analyzed within the same source-zero versus
Green-pole framework. Recognizing these cancellations shifts attention from
structural order alone to the gauge-like operator structure of the fields
carried by matter. This perspective suggests new ways to classify and design
disordered systems by controlling not only where material is placed, but how
physical sources are generated from it. In this sense, disordered matter can
remain structurally noisy while being quiet in the physical variables that
control response.

\section{Materials and Methods}

\subsection{Numerical calculations}
All numerical examples were computed on two-dimensional periodic square
domains with side length $L$, discretized by $N\times N$ grid points. We set
$\Delta x=L/N=1$, so $L$ and $N$ are numerically equal; throughout the text,
$N$ denotes grid resolution, not the number of inclusions. The main spatial
maps and spectral panels use $N=512$, while finite-size comparisons use
$N=128,256,512$, and $1024$. Derivatives were evaluated spectrally,
$\partial_x\rightarrow i k_x$ and $\partial_y\rightarrow i k_y$, and the
zero mode was removed from both source and response calculations. The three
source fields were generated from ordinary short-range-disordered parent
fields by the corresponding local operators:
$\eta=\operatorname{inc}\varepsilon^*$ for eigenstrain,
$\rho_b=-\nabla\cdot\mathbf P$ for polarization, and
$J_b=(\nabla\times\mathbf M)_z$ for magnetization.

The responses were computed in Fourier space using the periodic Green
operators described in the main text. For elasticity, the residual stress was
obtained from the Airy representation
$\Delta^2\chi=Y\eta$ and
$\sigma_{xx}=\partial_{yy}\chi$,
$\sigma_{yy}=\partial_{xx}\chi$,
$\sigma_{xy}=-\partial_{xy}\chi$. For electrostatics, we solved
$-\Delta\phi=\rho_b$ with $\mathbf E=-\nabla\phi$. For magnetostatics, we
solved $-\Delta A_z=\mu_0J_b$ with
$\mathbf B=\nabla\times(A_z\mathbf e_z)$.

Spectral densities $\tilde{\chi}_X(k)$ were obtained by radial shell averaging
of squared Fourier amplitudes, using the vector or tensor amplitude when
appropriate. Low-wavenumber exponents were fitted from
$\log\tilde{\chi}_X(k)=\beta_X\log k+C$ within pre-specified low-$k$ windows;
the selected windows, shell counts, and bootstrap standard errors are reported
in the \textit{SI Appendix}. To isolate the role of source architecture, each
physical source was compared with an independently generated white source
normalized realization by realization to have the same rms amplitude, and both
sources were passed through the same response operator.

\section*{Data Availability}
The codes and data are available upon request.

%\acknow{This work was supported by the Army Research Office under Cooperative Agreement Number W911NF-22-2-0103.}

%\showacknow{} % Display the acknowledgments section

\section*{Author Contributions}
Author contributions: L.Z., H. W. and Y.J. designed research; L.Z., H.W. and Y.J. performed research; L.Z., H.W. and Y.J. wrote the paper; L. Z., H.W., and Y.J. edited the paper.

\section*{Competing Interests}
The authors declare no competing interest.

% Bibliography
\bibliography{field_hu_references, network}

\end{document}